\documentclass[apjl]{emulateapj}
\usepackage{pstricks}
\usepackage{apjfonts}
\def\chandra{{\em Chandra}\/}
\def\einstein{{\em Einstein}\/}
\def\bi{\bfseries\itshape}
\def\hseventy   {$H_0$=70~km$\;$s$^{-1}\,$Mpc$^{-1}$}

\begin{document} 

\title{{\em CHANDRA}\/ OBSERVATION OF THE MERGING CLUSTER A168: A LATE STAGE
IN THE EVOLUTION OF A COLD FRONT}

\author{Eric J. Hallman\altaffilmark{1} 
and Maxim Markevitch\altaffilmark{2,3}} 

\altaffiltext{1}{Department of Astronomy, University of
Minnesota, 116 Church St. SE, Minneapolis, MN 55455, hallman@astro.umn.edu}

\altaffiltext{2}{Harvard-Smithsonian Center for Astrophysics, 60 Garden Street,
Cambridge, MA 02138, maxim@head.cfa.harvard.edu} 

\altaffiltext{3}{Space Research Institute, Profsoyuznaya 84/32, Moscow
117997, Russia}

\setcounter{footnote}{3}

\begin{abstract}

We present \chandra\ observations of the cool cluster A168, for which
previous X-ray imaging and optical studies indicated a merger of two
subclusters nearly in the plane of the sky.  We derive a temperature map for
A168, which shows that the merger has proceeded beyond the core passage and
is near subcluster turnaround. It also reveals an unusual feature --- the
gas core of one of the subclusters forms a tongue-like structure extending
ahead (in the direction of motion) of the subcluster center. The coolest
cluster gas is found in a crescent-shaped region at the tip of this tongue,
and forms a cold front in pressure equilibrium with the external gas.  In
contrast with this feature's forward location, previously observed
merger cold fronts (e.g., A3667, 1E0657--56) lagged behind their host
subclusters, as expected in the presense of ram pressure. We propose that
A168 illustrates a much later stage in the evolution of a cold front, when
its host subcluster approaches the apocenter of the merger orbit where the
ram pressure on its gas drops sharply.  As a result, a large chunk of
the subcluster gas ``slingshots'' past the dark matter center, becomes
unbound from the subcluster and expands adiabatically, as seen in some
recent hydrodynamic simulations.

\end{abstract}

\keywords{Galaxies: clusters: individual (A168) --- intergalactic medium ---
X-rays: galaxies}

\section{Introduction}

Galaxy clusters form via infall and merger of larger subunits. Such
mergers generate a wealth of transient hydrodynamic phenomena in the
intracluster gas that can be studied in X-rays, most notably shocks
(e.g., Schindler \& M\"uller 1993; Burns 1998; Ricker \& Sarazin 2001).  In
addition, typical clusters have dense gas cores that appear to be able to
survive a merger and form the prominent contact discontinuities, or ``cold
fronts'', revealed in \chandra\ high-resolution X-ray images of many merging
and relaxed clusters (e.g., Markevitch et al.\ 2000; Vikhlinin, Markevitch
\& Murray 2001b; Markevitch, Vikhlinin, \& Mazzotta 2001; Kempner, Sarazin,
\& Ricker 2002). Cold fronts are seen as sharp edges in the cluster X-ray
images and temperature maps.  The brighter side of the jump has a lower gas
temperature than the faint side by a ratio similar to the density jump,
indicating rough pressure balance across the front. This distinguishes cold
fronts from the less frequently observed bow shock discontinuities in which
the hotter side is also denser. Cold fronts are interpreted as a boundary of
a dense subcluster core moving through a hotter ambient medium (Markevitch
et al.\ 2000), though the term is often used to refer to the entire cold
region.  This phenomenon provides novel tools to study the cluster physics.
For example, cold fronts are very sharp, with a width smaller than the
electron Coulomb mean free path, which places severe upper limits on
electron diffusion and thermal conduction across the front (Vikhlinin et
al.\ 2001b; see also Ettori \& Fabian 2000).  A cold front in A3667 also
indicates a parallel magnetic field is necessary to stabilize it against the
onset of hydrodynamic instabilities (Vikhlinin et al.\ 2001a).  Merger cold
fronts have been reproduced in recent hydrodynamic simulations (e.g.,
Bialek, Evrard, \& Mohr 2002; Nagai \& Kravtsov 2003; Mathis et al.\ 2003),
although with limited spatial resolution.

This paper analyzes \chandra\ observations of A168 ($z=0.045$).  Earlier
X-ray imaging of A168 with \einstein\ IPC revealed irregular structure
indicative of a merger \citep{ulmer,jf99}.  The ends of its extended X-ray
structure coincide with two galaxy subclusters, the northern of which is
dominated by a cD galaxy UGC 00797 and has few other members, while the rich
southern subcluster has no dominant galaxy (Fig.\ 1{\em a}). More evidence
of a merger is provided by a recent detailed optical study which shows that
the redshift distributions of the two subclusters differ significantly
\citep{yang}.  This interesting merger, apparently occurring near the plane
of the sky, is the subject of this \chandra\ study.  We use \hseventy\ and
quote 90\% confidence intervals throughout the paper.

\begin{figure*}
\pspicture(0,16.3)(18.0,24.0)

\rput[tl]{0}(0.5,24.0){\includegraphics[bb=70 100 600 490,clip,height=14.0cm]%
{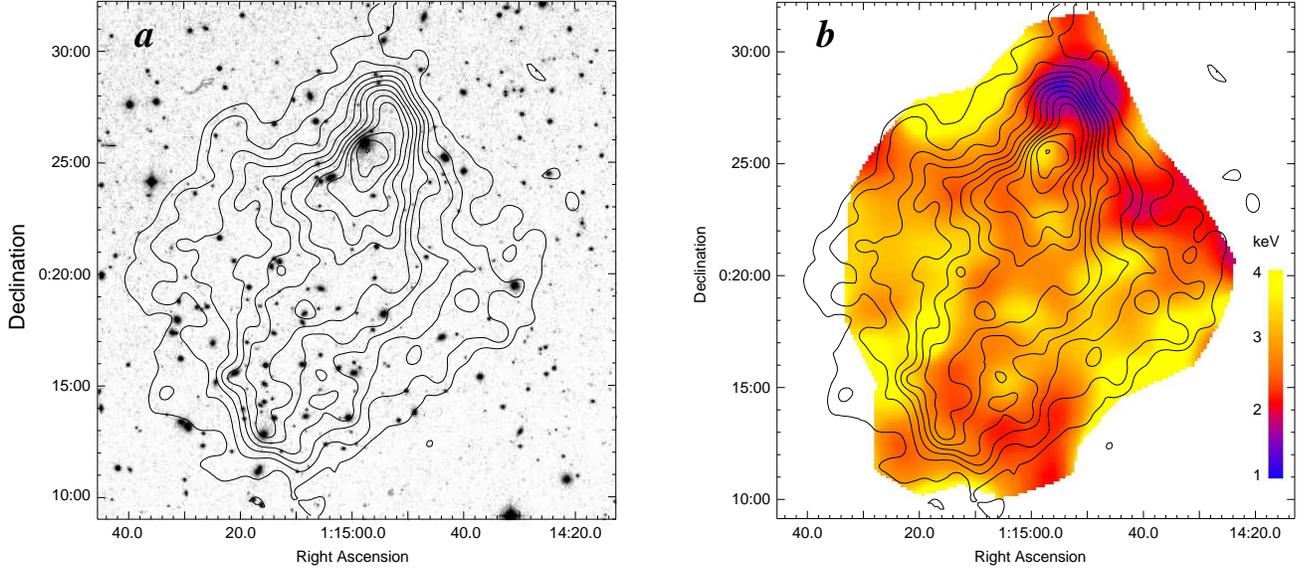}}

\endpspicture

\caption{({\em a}\/) Contours of 0.8--6.0 keV \chandra\ X-ray brightness 
  (linear spacing; image smoothed by $\sigma=24''$), overlaid on Palomar
  Digitized Sky Survey optical image of A168 field. ({\em b}\/) Projected
  temperature map (colors) overlaid with the image contours.}
\end{figure*}

\section{Data Analysis}

Two ACIS-I observations of A168 were done in 2002 January and November for
40 ks and 37 ks, respectively. The first pointing covered the northern
region of the cluster, including the cD galaxy, and the second covered the
southern X-ray elongation and overlapped with the first pointing.  Our
analysis procedure largely follows Markevitch \& Vikhlinin (2001).
The background was modeled using blank-sky files from the \chandra\
calibration database.
A correction for the low-energy absorption
caused by the ACIS contamination buildup (Plucinsky et al.\ 2003)
was included in the ARFs.  
Point sources were masked out. The two offset pointings were co-added; the
resulting combined ACIS image is overlaid on the DSS plate in Fig.\ 1{\em
a}.

To study the global properties of the cluster, we extracted a 0.8--6 keV
spectrum from a $r=10'$ circle around the surface brightness centroid
($\alpha$=01:15:01, $\delta$=+00:20:26)
Using the Galactic absorption column $N_H=3.48\times 10^{20}$ cm$^{-2}$, we
obtained $T_e=2.8 \pm 0.1$ keV, and a chemical abundance of $0.30 \pm 0.05$
relative to solar. This temperature is in good agreement with the
earlier \einstein\ MPC measurement of $2.6^{+1.1}_{-0.6}$ keV
\citep{david}. However, as is seen below, the cluster is highly
non-isothermal.

\subsection{Temperature Map}

We derived a projected temperature map by creating cluster images in 5
energy bands, 0.8--1.3--2--3--4--6 keV. These images were each adaptively
smoothed by a Gaussian in the $\sigma=20''-60''$ interval using an identical
smoothing pattern.  The flux values for each energy at each image pixel
were fit by a MEKAL model with absorption set to the Galactic value and the
chemical abundance set to the overall cluster value. The resulting
map is shown in Fig.\ 1{\em b}, overlaid on the brightness contours.  To
check the significance of the temperature variations, and to look for
abundance variations, we also extracted spectra from 12 regions of the image
(Fig.\ 2) and performed the usual spectral fitting with either fixed or free
abundances. No significant abundance variations are detected. Figure 2 shows
the fits with fixed abundances; there is good agreement with the smooth
temperature map.

\begin{figure}[b]
\pspicture(0,4.8)(9.0,16.2)

\rput[tl]{0}(0.0,16.4){\includegraphics[bb=70 20 510 540,clip,height=18.8cm]%
{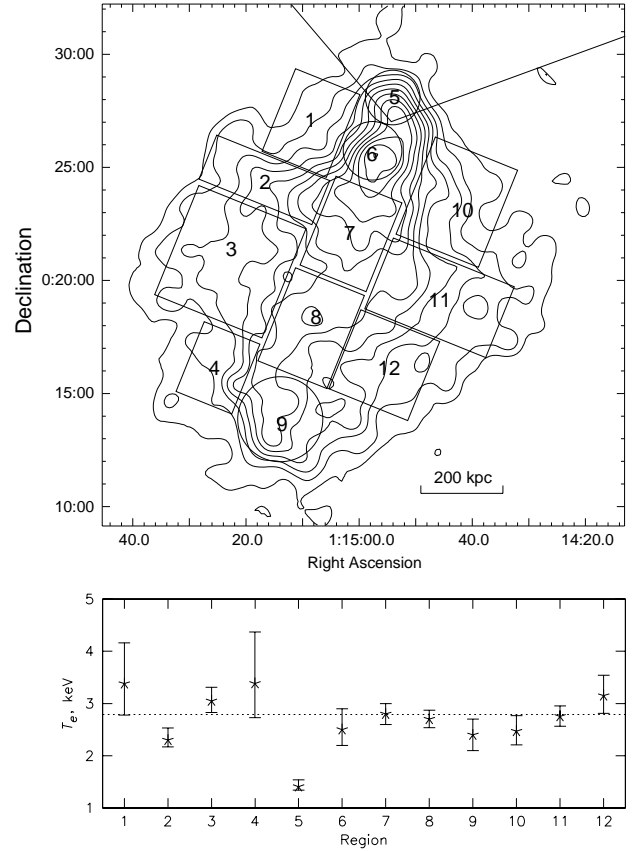}}

\endpspicture

\caption{X-ray brightness contours overlaid with numbered spectral
  extraction regions.  Underneath is plot of fitted temperature from spectra
  extracted from these regions. Dotted line is global cluster temperature
  fit. The two angled lines originating in region 5 are the sector border
  for the annular regions used for radial profiles.}

\end{figure}

\begin{figure*}
\pspicture(0,12.6)(18,23.2)

\rput[bl]{0}(1.5,12.3){%
\includegraphics[scale=0.7]{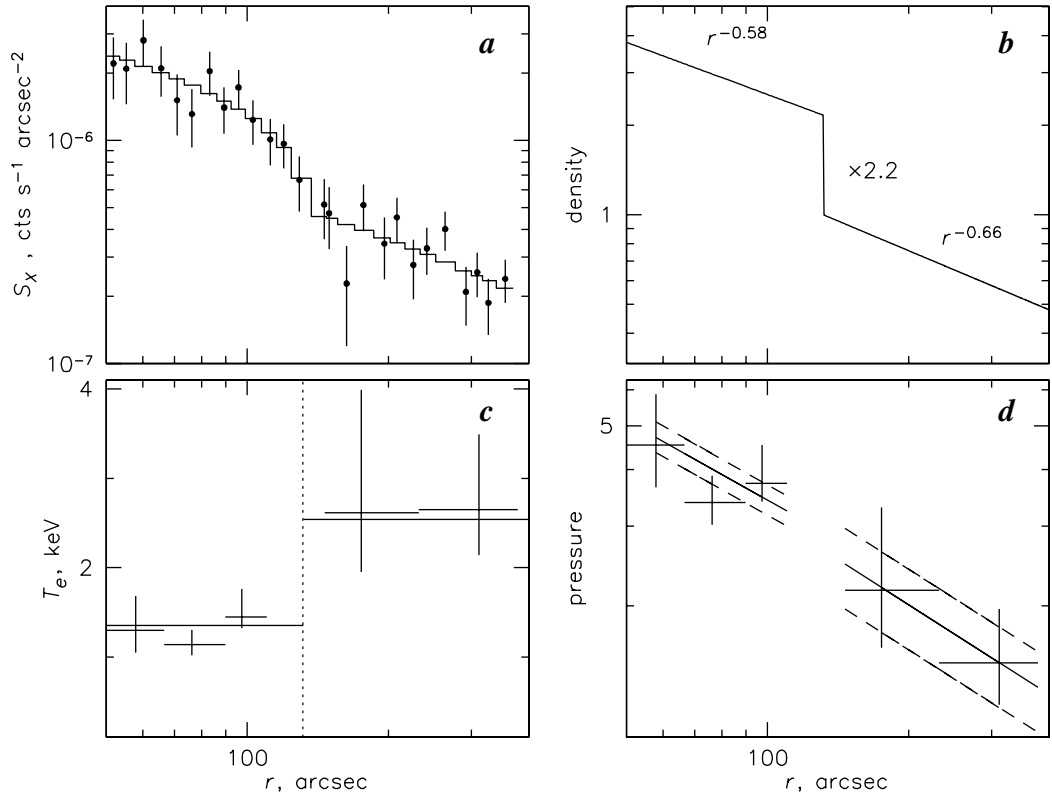}}

\rput[bl]{0}(7.8,22.7){\bi\large a}
\rput[bl]{0}(14.7,22.7){\bi\large b}
\rput[bl]{0}(7.8,17.8){\bi\large c}
\rput[bl]{0}(14.7,17.8){\bi\large d}

\endpspicture

\caption{({\em a}\/) 0.8--3.0 keV X-ray brightness profile across the
  northern front. Regions extracted for this profile use concentric annular
  sectors cut at the lines indicated in Fig.\ 2. Solid line is the
  brightness profile resulting from the projected best-fit spherical density
  model shown in ({\em b}\/).  Density is in arbitrary units. ({\em
  c}\/)  Temperature profile across the same front. Solid lines are
  the combined single temperature fits for all interior and all exterior
  regions, respectively. Vertical line indicates location of best-fit
  density jump. ({\em d}\/) Pressure profile generated from multiplication
  of the density model with the fitted spectral temperatures at each
  point. Solid lines indicate the pressure from the combined temperature fits
  inside and outside the jump, dashed lines are the errors for those. Units
  are arbitrary.}
\label{profplt}
\end{figure*}

The most striking feature in the temperature map and the X-ray image is the
cool crescent-shaped region in the north.  It does not coincide with the gas
density peak approximately at the position of the cD galaxy, which would
make it similar to many cooling flow clusters. Rather, it lies at the tip
of the prominent tongue-like X-ray brightness feature extending 200 kpc
north of the cD and ending with an apparent cold front. We will discuss
this ``northern front'' in more detail below (\S\ref{sec:front}).

Other interesting features in the temperature map include a
cool region at the southern subcluster (region 9), a cool arm-like
eastward X-ray extension (region 2), a symmetric cool feature less prominent
in the X-ray image (region 10), and hotter areas on either side of the
central brightness ridge (regions 3, 12 and part of 11). An important
thing to note is that the temperature map {\em does not}\/ show a
characteristic hot region between the merging subclusters, such as we would
expect if they were approaching each other at present (e.g., Ricker \&
Sarazin 2001; Belsole et al.\ 2003).

\section{Dynamical State of A168}

In a recent study of the A168 member redshifts, Yang et al.\ (2004) show
that the galaxies near the northern X-ray peak and the southern X-ray
elongation separate into statistically distinct groups.  They conclude
that the northern and southern galaxy subgroups are on a collision course
roughly in the plane of the sky.  However, since the line of sight
orientation of the subclusters is unknown, the determination of the
direction of their relative motion from the redshift data alone is
ambiguous, so the Yang et al.\ scenario for A168 is not unique. 

The new X-ray data suggest a more likely scenario for the merger stage and
geometry.  
The presence of the ``northern front'' indicates a motion of the northern
subcluster in the northern direction (i.e., away from the other
subcluster), either at present or very recently. An interpretation which
better fits this observation, as well as the lack of shock-heated gas
between the two X-ray peaks in our temperature map, is that the A168
subclusters have already passed one another at least once. Simulations,
e.g., by \citet{rick} and \citet{mathis}, support this
interpretation. Immediately after one pass of subclusters, they predict an
elongated appearance of the X-ray surface brightness and a ridge of cooler
gas between the subcluster centers. In addition, \citet{mathis}
predict the qualitative features of the northern front very well; this
is further discussed in \S\ref{sec:front}.

In this late merger scenario, the marginally cooler southern subcluster
(region 9 in Fig.\ 2) is an analog to the northern front, although not quite
as prominent.  The morphology of the temperature structure is qualitatively
similar, in that the cold gas is displaced toward the tip of the X-ray
elongation. The slightly hotter gas on the two sides of the cluster axis
(regions 3 and 12) could be what remains of the shock-heated gas between the
colliding subclusters that has had time to expand adiabatically as those
subclusters have moved apart.  The cool eastern arm in region 2 may
have analogs in A3667, where a similar feature was interpreted as a
large-scale eddy created by Kelvin-Helmholtz instability (Mazzotta et al.\
2002), and in simulations by Ricker \& Sarazin.

\section{The Northern Front}
\label{sec:front}

To study the pressure across the northern brightness edge, we extract a
surface brightness and temperature profile from concentric elliptical
annular regions (with the ellipse selected to follow the brightness edge)
inside and outside the edge, confined to the sector between the two
lines on Fig.\ 2. The 0.8--3.0 keV surface brightness in this sector is
plotted in Fig.\ 3{\em a}. The solid line is a projection of the best-fit
spherically symmetric (in the interesting region of space) density model
that consists of two power-law radial profiles and an abrupt jump at a
certain radius (Fig.\ 3{\em b}), all of which were free parameters.  To
compare the model to the brightness profile, we projected it and took into
account the different emissivity due to the accompanying temperature change
(see below). The best-fit density jump is a factor of $2.2^{+0.6}_{-0.4}$.
The brightness jump is not as sharp as in some other clusters, with a
transition region of roughly 30\arcsec\ or 27 kpc, a factor of a few wider
than the upper limit on the width of the unresolved front in A3667
(Vikhlinin et al.\ 2001b).  A possible explanation is a small nonzero
angle of the subcluster velocity from the sky plane (e.g., Mazzotta et al.\
2001). 

A temperature profile across the edge is derived in wider annular regions in
order to include sufficient counts. Since the edge is not particularly
sharp, the inner and outer spectral regions are separated by a 30\arcsec\
gap to avoid contamination of the outer regions by contribution
from the brighter inner region.
We performed an approximate deprojection of the contribution of the outer,
hotter spherical regions to the inner spectra, using the density model
fitted above.  This resulted in a small temperature correction for the bins
nearest to the edge. Figure 3{\em c}\/ shows the resulting temperature
profile.  Also shown are a simultaneous fit for the spectral regions inside
the edge, $T=1.35\pm0.1$ keV, and for the outer regions,
$T=2.5^{+0.7}_{-0.4}$ keV. The individual bins are consistent with these
averages on both sides of the jump. The ratio of these temperatures is
$1.9^{+0.5}_{-0.3}$, the reverse of the density jump obtained above,
which clearly identifies this feature as a cold front.

The temperatures and the density model provide a pressure profile across
the front (Fig.\ 3{\em d}), which shows that thermal pressure is continuous
within the errors. This indicates that the present velocity of the cool
gas relative to the ambient gas is consistent with zero, with a 90\% upper
limit on the Mach number of $M<0.7$ assuming a stationary flow (e.g.,
Vikhlinin et al.\ 2001a).

What makes the cold front in A168 unusual is that the gas is apparently
not driven by, or lagging behind, the associated galaxy subcluster (Fig.\
1{\em a}). This is in contrast to 1E0657--56 (Markevitch et al.\ 2002; Clowe
et al.\ 2004) and A3667 (Vikhlinin \& Markevitch 2002), the two cold fronts
where the subcluster mass centroid could be located.  The gas is expected to
lag because, unlike dark matter, it is affected by the ram pressure of the
ambient gas.  Instead, the cool gas in A168 lies ahead of the cD galaxy that
should be the center of the local gravitational potential.

The development of fronts like the one in A168 is predicted by simulations
(most clearly those by Mathis et al.\ 2003) when a cool dense subcluster
emerges from the collision site and is near its apocenter, but no clear
physical explanation for such detachment was proposed. The simple
explanation which we propose is one in which the ram pressure from the
motion of the subcluster through the ambient medium initially pushes the gas backwards from the dark matter core centered on the cD galaxy, and it
trails the dark matter as in other clusters. However, once the
subcluster approaches the apocenter of its orbit, the ram pressure drops
sharply due to the combination of reduced ambient gas density and lower
subcluster velocity.  At this point, the gas which was trailing the dark
matter core ``slingshots'' past it (driven in part by the gas internal
pressure and in part by the subcluster's gravity) into the still lower
density medium, while the underlying dark matter core turns around.  The
leading part of the gas core becomes unbound from the subcluster, expands,
cools adiabatically and forms the cold tip we observe in A168.

Heinz et al.\ (2003) predicted that the motion of a subcluster through a
cluster medium should generate gas flows inside the subcluster which
transport gas to its front edge, which alone should produce a
crescent-shaped cool region near the front, just as observed in A3667 (Heinz
et al.) and in our cluster.  A combination of this process and the ram
pressure ``slingshot'' proposed above fully explains the structure and
location of the northern front in A168.

The A168 front resembles a cool-tipped gas tongue in A754, which was
proposed to be the result of the gas ``sloshing'' out of a subcluster at an
advanced merger stage \citep{mark03}.  Unlike the messy A754, A168 is simple
and shows this process more clearly.

An interesting implication of our A168 observation is that it shows how a
large chunk of the cool subcluster gas unbinds from the subcluster and
deposits itself into the ambient gas.  Since it is in pressure equilibrium
and has zero velocity relative to the ambient gas, it will persist there for
a long time as a distinct region (e.g., Mathis et al.\ 2003). This shows
that a merger is inefficient in mixing the different gas phases.

\section{Summary}

We have analyzed \chandra\ observations of a merging cluster A168. The
cluster brightness and projected temperature distribution suggest that the
two subcluster cores have passed one another and are in the process of
turning around. The coolest cluster gas is found at the tip of an
interesting tongue-like structure, which extends ahead of the northern
subcluster (i.e., away from the merger center) and ends with a cold front.
This cold front differs from those observed earlier (e.g., in A3667 and
1E0657--56) which lag behind their host subclusters due to the ram pressure
effects.  A168 presents the first example of a cold front at a late stage of
its evolution, when the subcluster reaches its apocenter, the ram pressure
of the ambient gas sharply decreases and the leading part of the subcluster
gas core ``slingshots'' ahead of its host dark matter concentration,
expanding adiabatically and forming an offset cool region.  According to
simulations (Mathis et al.\ 2003), this cool gas will remain there as a
distinct region as the dark matter subcluster falls back.

\acknowledgments

We thank Alexey Vikhlinin for his spectral analysis software and Bill Forman
and Tom Jones for valuable comments.  Support was provided by NASA contract
NAS8-39073, \chandra\ grant GO2-3165X, the NSF grant AST03-07600, and the
Minnesota Supercomputing Institute.


\end{document}